\documentclass[lettersize,journal]{IEEEtran}
\usepackage{amsmath,amsfonts}
\usepackage{algorithmic}
\usepackage{algorithm}
\usepackage{array}
\usepackage[caption=false,font=normalsize,labelfont=sf,textfont=sf]{subfig}
\usepackage{textcomp}
\usepackage{stfloats}
\usepackage{url}
\usepackage{verbatim}
\usepackage{graphicx}
\usepackage{amssymb}
\usepackage{cite}
\usepackage{booktabs}
\usepackage{pifont} 
\usepackage{hyperref}
\usepackage{adjustbox}
\usepackage{tabularx, makecell, multirow} 
\usepackage[dvipsnames]{xcolor}

\begin{document}

\title{UniUltra: Interactive Parameter-Efficient SAM2 for Universal Ultrasound Segmentation}

\author{Yue Li, Qing Xu, Yixuan Zhang, Xiangjian He, \IEEEmembership{Senior Member, IEEE}, Qian Zhang, Yuan Yao, Fiseha B. Tesem, Xin Chen, \IEEEmembership{Senior Member, IEEE}, Ruili Wang, Zhen Chen,  Chang Wen Chen, \IEEEmembership{Fellow, IEEE}  
	\thanks{\quad This work is partially supported by the Yongjiang Technology Innovation Project (2022A-097-G), Zhejiang Department of Transportation General Research and Development Project (2024039), and National Natural Science Foundation of China grant (UNNC: B0166). \textit{(Equal contribution: Yue Li and Qing Xu, Corresponding authors: Xiangjian He, Qian Zhang)}}
    \thanks{\quad Y. Li is with School of Computer Science, University of Nottingham Ningbo China, Ningbo, Zhejiang, China, and with School of Computer Science, University of Nottingham, UK (e-mail: yue.li3@nottingham.edu.cn).}
    \thanks{\quad Q. Xu, Y. Zhang, X. He, Q. Zhang, Y Yao, F. B. Tesem, are with School of Computer Science, University of Nottingham Ningbo China, Ningbo, Zhejiang, China (e-mail: sean.he@nottingham.edu.cn).}
    \thanks{\quad X. Chen is with School of Computer Science, University of Nottingham, UK (e-mail: xin.chen@nottingham.ac.uk).}
    \thanks{\quad R. Wang is with the School of Mathematical and Computational Sciences, Massey University, Auckland 0632, New Zealand, also with the School of Computer Science, University of Nottingham Ningbo China, Ningbo 315104, China, and also with the School of Data Science and Artificial Intelligence, Wenzhou University of Technology, Wenzhou, China (e-mail: ruili.wang@massey.ac.nz)}
    \thanks{\quad Z. Chen is with Hong Kong Institute of Science \& Innovation, Chinese Academy of Sciences, Hong Kong SAR (e-mail: zchen.francis@gmail.com).}
    \thanks{\quad C. W. Chen is with The Hong Kong Polytechnic University, Hong Kong (e-mail: changwen.chen@polyu.edu.hk).}
    }


\markboth{IEEE Transactions on Multimedia}%
{}



\maketitle

\begin{abstract}
The Segment Anything Model 2 (SAM2) demonstrates remarkable universal segmentation capabilities on natural images. However, its performance on ultrasound images is significantly degraded due to domain disparities. This limitation raises two critical challenges: how to efficiently adapt SAM2 to ultrasound imaging while maintaining parameter efficiency, and how to deploy the adapted model effectively in resource-constrained clinical environments. To address these issues, we propose UniUltra for universal ultrasound segmentation. Specifically, we first introduce a novel context-edge hybrid adapter (CH-Adapter) that enhances fine-grained perception across diverse ultrasound imaging modalities while achieving parameter-efficient fine-tuning. To further improve clinical applicability, we develop a deep-supervised knowledge distillation (DSKD) technique that transfers knowledge from the large image encoder of the fine-tuned SAM2 to a super lightweight encoder, substantially reducing computational requirements without compromising performance. Extensive experiments demonstrate that UniUltra outperforms state-of-the-arts with superior generalization capabilities. Notably, our framework achieves competitive performance using only 8.91\% of SAM2's parameters during fine-tuning, and the final compressed model reduces the parameter count by 94.08\% compared to the original SAM2, making it highly suitable for practical clinical deployment. The source code is available at \url{https://github.com/xq141839/UniUltra}.
\end{abstract}


\section{Introduction}
\label{sec:introduction}

\IEEEPARstart{U}{ltrasound} imaging is a non-invasive, real-time diagnostic tool that plays an important role in modern medical practice \cite{shu2022cross, song2024centerformer}. Its ability to provide dynamic and high-resolution soft-tissue images makes it indispensable in a variety of clinical applications, ranging from prenatal care \cite{peahl2023routine} to cardiac assessment. Ultrasound image segmentation is a critical task in computer vision, and precise segmentation of lesions and organs is a key step in disease diagnosis and treatment planning \cite{eswaran2023applying}. However, inherent challenges in ultrasound images, such as speckle noise, low contrast and artifacts, pose significant difficulties for accurate segmentation. In addition to these challenges, similar intensity distributions between the target object and the background \cite{chen2022aau}, blurred boundaries and variability of the target object also add to the difficulty of accurate ultrasound image segmentation.

\begin{figure}[!t]
  \centering
  \includegraphics[width=\linewidth]{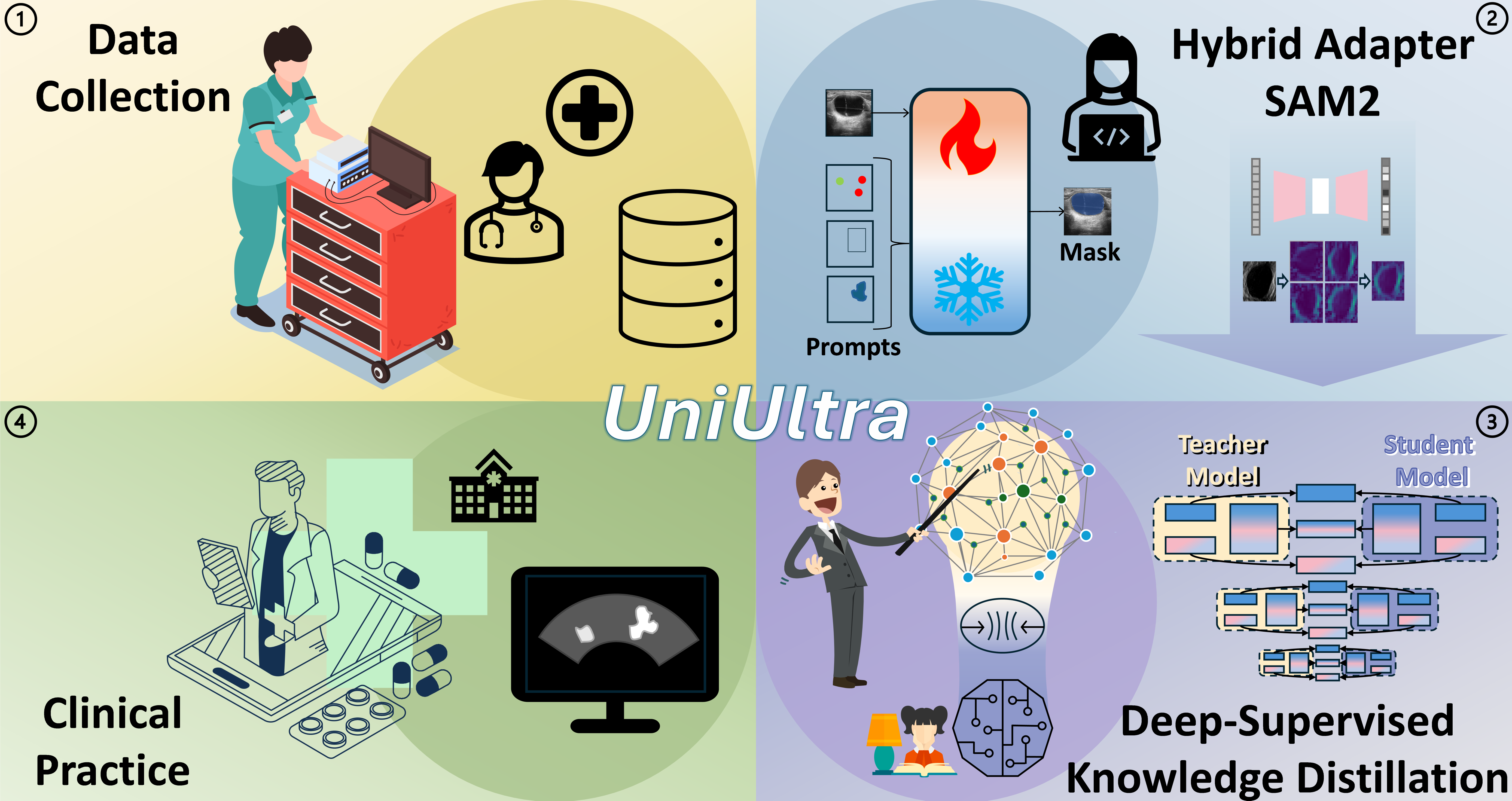}
  \caption{The design pipeline of UniUltra. The process begins with data collection, where sonographers acquire and label ultrasound images. Subsequently, the SAM2 model undergoes fine-tuning on the labeled ultrasound data using PEFT-based CH-Adapters with additional touch input for interaction. Finally, DSKD is employed to compress the size of the fine-tuned SAM2, ensuring its universality and real-world clinical applicability.}
  \label{fig:1}
\end{figure}

\begin{figure*}[!t]
  \centering
  \includegraphics[width=1\linewidth]{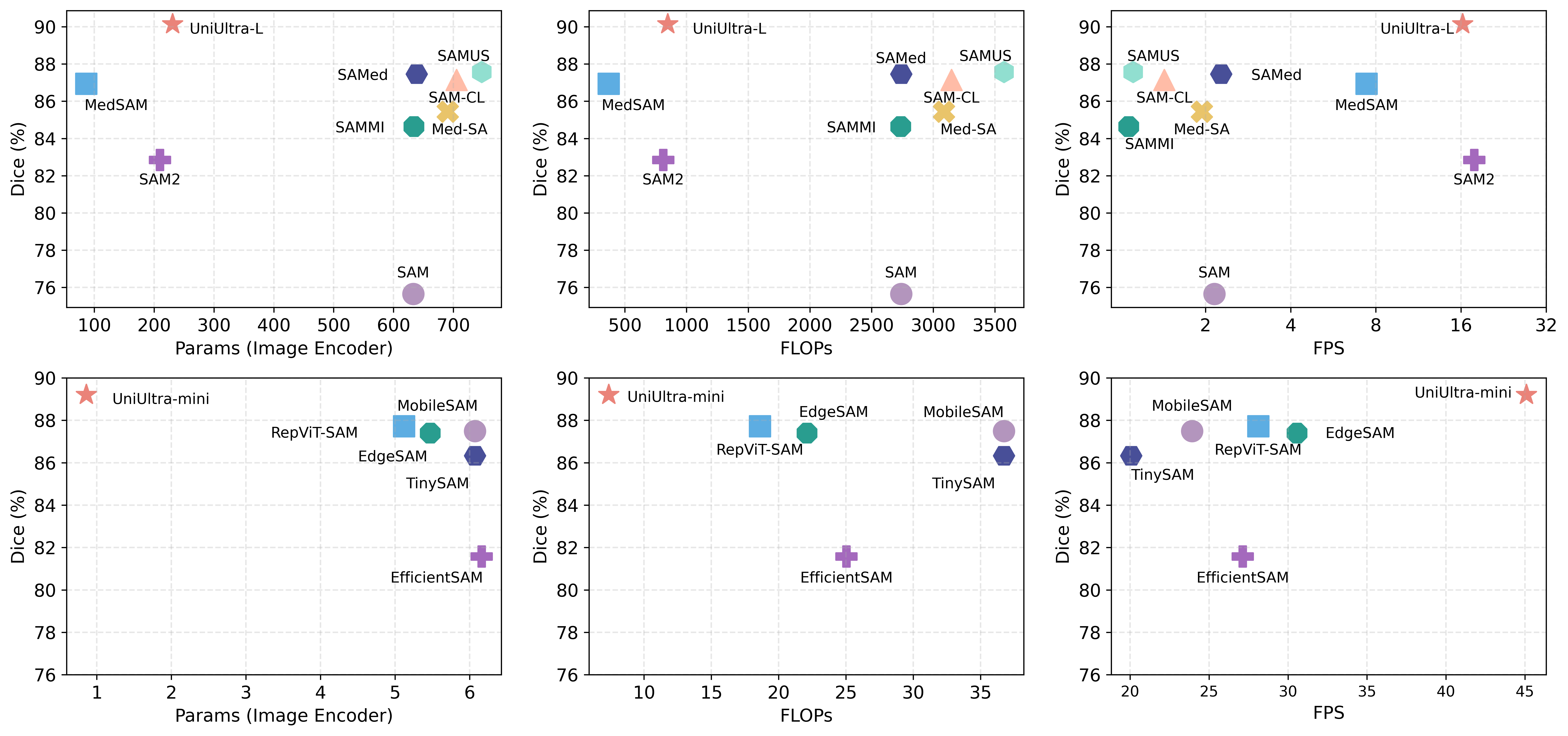}
  \caption{Performance and efficiency comparisons between state-of-the-arts and our UniUltra. Results demonstrate the superior universality of UniUltra across different ultrasound scenarios. In particular, the image encoder of UniUltra-mini uses only 0.86M parameters, revealing remarkable memory efficiency.}
  \label{fig:params}
\end{figure*}

In past studies, UNet \cite{ronneberger2015u} was proposed for medical image segmentation, which had a symmetric U-shaped architecture and an encoder-decoder structure that allowed the fusion of low-level and high-level features through skip connections. UNet and its variants \cite{isensee2021nnu,chen2024transunet,rahman2024emcad,nam2024modality} have become cornerstones due to their effectiveness in capturing contextual information and local details. AAU-Net \cite{chen2022aau} enhanced the effectiveness of feature fusion through adaptive aggregation units and had superior performance on ultrasound of breast lesion segmentation. However, the generalization of the above models across different datasets has always been a concern. To solve the above problems, Segment Anything Model (SAM) \cite{kirillov2023segment} series used an interactive prompt to achieve generalization performance in diverse natural scenarios \cite{zhu2025exploiting, liu2025segmenting, zheng2024black, lan2023foodsam, mei2025rog, li2024low}. The latest version, SAM2 \cite{ravi2024sam}, was a fast segmentation technique that could generalize to unfamiliar objects and images with zero-shot generalization and performed well in natural image segmentation. However, when applied to medical images directly, its effectiveness was greatly limited due to the significant domain differences between natural and medical images. The researchers then proposed MedSAM \cite{ma2024segment}, a SAM variant tailored for medical images, to bridge the domain gap. However, both SAM and its variants were constrained by their large parameter sizes, making it challenging to deploy them in clinical practice. Although existing methods \cite{wu2024trans} could significantly reduce the number of trainable parameters when fine-tuning, the final size of the model was still large.

To address the aforementioned challenges, we propose UniUltra, a universal ultrasound segmentation framework that simultaneously tackles adaptation efficiency and clinical deployment requirements. As illustrated in Figure \ref{fig:1}, we first devise
a novel context-edge hybrid adapter (CH-Adapter) that employs parameter-efficient fine-tuning (PEFT) by integrating context-aware and edge-aware components into the SAM2 image encoder. The context-aware component captures global semantic information, while the edge-aware component leverages four-directional Sobel filters to enhance fine-grained boundary perception. This design enables effective domain adaptation using only 8.91\% of SAM2's parameters during fine-tuning. Then, we design a deep-supervised knowledge distillation (DSKD) that transfers knowledge from the fine-tuned SAM2 to a lightweight architecture while preserving critical feature extraction capabilities. The resulting compressed model achieves a 94.08\% parameter reduction compared to the original SAM2. As demonstrated in Figure \ref{fig:params}, UniUltra-L and UniUltra-mini achieve superior performance-efficiency trade-offs compared to state-of-the-art (SOTA) methods, delivering high Dice scores with significantly fewer parameters, lower FLOPs, and higher FPS. This combination of accuracy and efficiency makes UniUltra particularly suitable for resource-constrained clinical environments.

In summary, our contributions are as follows:
\begin{itemize}
    \item We propose UniUltra, a new pipeline that adapts SAM2 from natural to ultrasound domains. UniUltra leverages touch interactions to achieve ultrasound segmentation with superior computation efficiency.
    \item We design a novel CH-Adapter that enables parameter-efficient fine-tuning of SAM2 for ultrasound imaging. The adapter integrates context-aware global perception with edge-aware boundary enhancement.
    \item We introduce the DSKD technique that provides stage-wise independent distillation for each encoder layer, achieving reduction of computation cost while maintaining representation capabilities for clinical deployment.
    \item Extensive experiments on four public datasets for internal validation and two additional datasets for external validation demonstrate UniUltra's superior computation, efficiency, and generalization capabilities across diverse ultrasound imaging scenarios.
\end{itemize}

\section{Related Work}

To explore the technological progress in ultrasound image segmentation in depth, this section first introduces methods for improving segmentation accuracy, robustness, and generalization ability. Then, the related research on efficient Segment Anything Models (SAMs) is reviewed.

\subsection{Ultrasound Image Segmentation}

In medical image segmentation, especially ultrasound, various deep-learning methods have been proposed to improve the accuracy and efficiency of segmentation. As a pioneering work, UNet \cite{ronneberger2015u} achieved remarkable results in medical image segmentation with its U-shaped structure and skip connection. The design of UNet allowed the model to pass information between the encoder and decoder, effectively combining contextual information and detailed features of the image. To further improve the performance of UNet, researchers proposed its variants \cite{chen2024transunet,isensee2021nnu,rahman2024emcad,nam2024modality,chen2022aau, xu2025mambavesselnet++, li2025hfa, chen2024mlmseg, qu2024eh, chen2024multi, luo2025lgffm}, attempting to improve the accuracy and robustness of segmentation by improving feature extraction and fusion mechanisms to capture more features under different receptive fields. However, their generalization on different datasets was a problem. Segment Anything Model (SAM) \cite{kirillov2023segment} achieved generalization performance by using interactive prompts. Although SAM, the latest version SAM2 \cite{ravi2024sam}, and its variants \cite{ke2024segment,zou2024segment} performed well in natural image segmentation, they faced the challenge of domain differences in medical image segmentation. Therefore, some SAM-based medical image segmentation models \cite{ma2024segment,huang2024segment} were proposed, which fully fine-tuned SAM on large public medical datasets collected by the researchers. However, the computational requirements, training costs, and large amounts of data costs inevitably increased. 

\subsection{Efficient Segment Anything Models}

The Methods based on parameter-efficient fine-tuning (PEFT) \cite{mangrulkar2022peft} made various improvements in SAM to fit specific tasks \cite{gowda2024cc, chen2025sam, lou2025nusegdg, xu2025co, zhang2025adapting}, avoiding the overfitting and high computational cost caused by full fine-tuning. Researchers inserted Adapters \cite{chen2023sam,chen2024sam2} into the image encoder of SAM to allow the model to learn task-specific knowledge and fine-tune SAM while only updating a few parameters. SAMed \cite{zhang2023customized}, Conv-LoRA \cite{zhong2024convolution}, and Med-SA \cite{wu2023medical} updated trainable SAM parameters via the Low-Rank Adaptation (LoRA) method \cite{hu2021lora}. Additionally, SAMUS \cite{lin2023samus} adapted SAM to ultrasound image segmentation by introducing parallel CNN branches. Despite using fewer trainable parameters, the final model size was still substantial, which limited the application in resource-constrained environments.

Knowledge Distillation (KD) \cite{hinton2015distilling} is a classic model compression method that guides lightweight student models to imitate teacher models with better performance and more complex structures \cite{xu2025hrmedseg}. Specifically, MobileSAM \cite{zhang2023faster} proposed decoupled distillation, using the lightweight image encoder to replace the original heavyweight one. Tinysam \cite{shu2023tinysam} was inspired by MobileSAM, replacing the image encoder with TinyViT \cite{wu2022tinyvit} and designing hard mining full-stage distillation strategy. EfficientSAM \cite{xiong2024efficientsam} leveraged masked image pre-training to compress the image encoder. EfficientViT-SAM \cite{zhang2024efficientvit} employed EfficientViT to replace the image encoder in SAM, achieving a significant speedup without performance loss through knowledge distillation and end-to-end training. RepViT-SAM \cite{wang2023repvit} further enhanced efficiency by integrating the RepViT model, optimized for mobile devices, into SAM, enabling nearly real-time inference on edge devices. De-LightSAM \cite{xu2025lightsam} introduced an end-to-end architecture for medical image segmentation, applying a Multi-Modal Decoupled Knowledge Distillation method to design a novel lightweight image encoder for producing discriminative visual features. However, existing KD methods did not fully consider the size of trainable parameters when compressing models, resulting in low training and inference efficiency in practical applications. We propose to consider using both fewer trainable parameters and compressing the model simultaneously.

\section{Methodology}

In this section, we introduce our proposed universal ultrasound segmentation framework, UniUltra. We first present the overall architecture of UniUltra. Then, our CH-Adapter and DSKD techniques are introduced. Finally, we provide the detailed training pipeline.

\begin{figure*}[!t]
  \centering
  \includegraphics[width=1\linewidth]{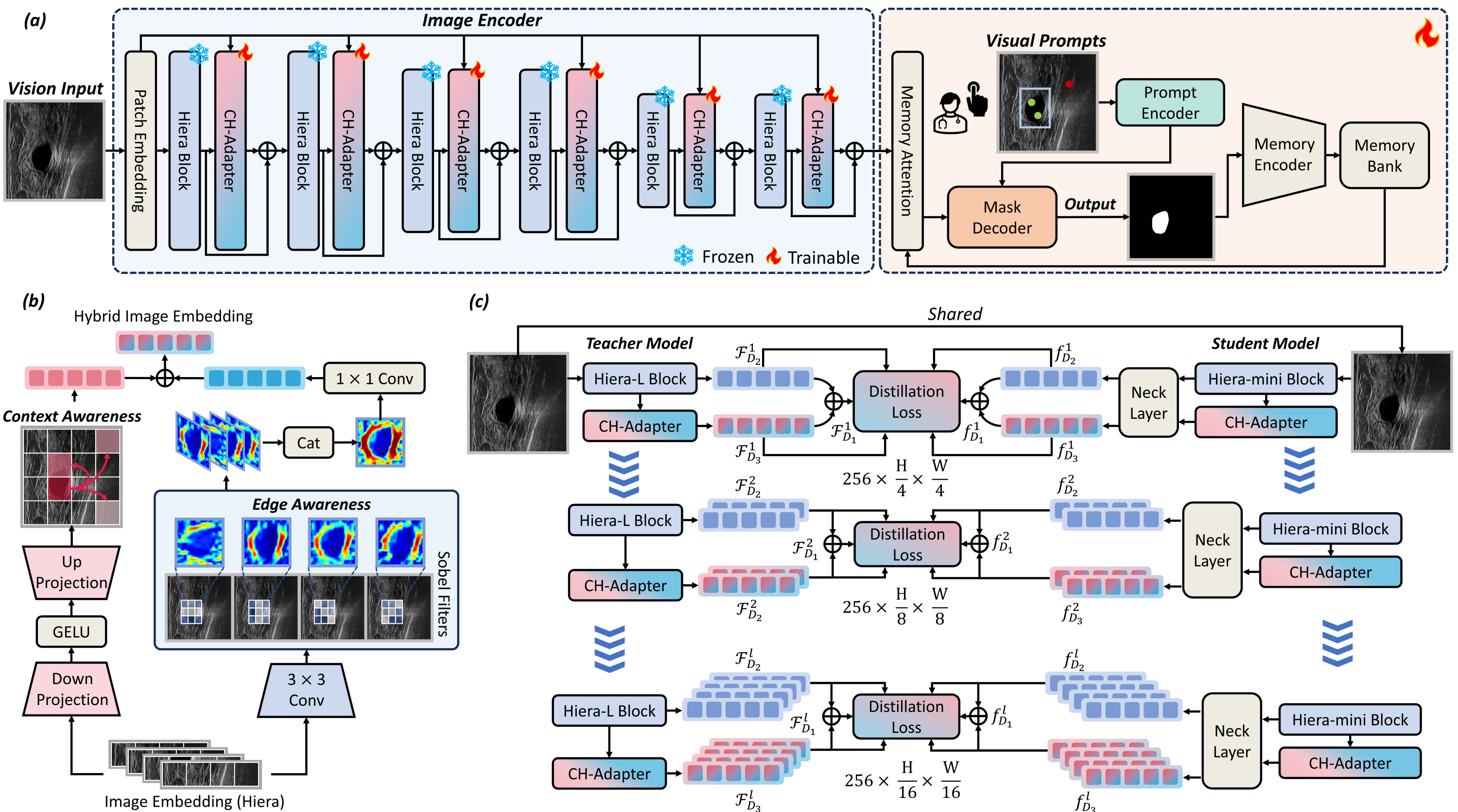}
  \caption{(a) The overview of the proposed UniUltra for universal ultrasound segmentation, consisting of (b) CH-Adapter and (c) DSKD. UniUltra provides a systematic pipeline that adapts SAM2 from natural to clinical deployment of ultrasound scenarios. Sonographers use bounding boxes on ultrasound equipment to outline target areas as touch input.}
  \label{fig:2}
\end{figure*}

\subsection{Overview of UniUltra}

As shown in Figure \ref{fig:2}(a), we present UniUltra to provide universal ultrasound segmentation with superior performance across diverse ultrasound imaging modalities. Given ultrasound images from various clinical scenarios, we first utilize the CH-Adapter to generate domain-adaptive image embeddings while maintaining parameter efficiency. These embeddings are then processed by the original decoding workflow of SAM2 to accurately predict segmentation masks. Following this, the DSKD technique transfers knowledge from the adapted model to a tiny architecture for clinical deployment.

In general, our UniUltra, comprising the CH-Adapter enhanced encoder and the original SAM2 decoding components, is specifically designed to tackle the challenges of domain adaptation and computational efficiency in ultrasound image segmentation. Meanwhile, the DSKD module eliminates the computational burden typically associated with large-scale models while preserving critical feature extraction capabilities. Therefore, these key modules enhance the segmentation capability of our UniUltra across a wide array of ultrasound images while ensuring clinical applicability.

\subsection{Context-Edge Hybrid Adapter}

The segmentation foundation models (e.g., SAM2 \cite{chen2024sam2}) demonstrate remarkable segmentation performance on natural images. However, direct application to ultrasound images suffers from significant domain disparities. Unlike natural images, ultrasound images exhibit unique characteristics, including speckle noise, acoustic shadows, and ambiguous tissue boundaries, leading to substantial performance degradation. Accurate boundary detection is particularly critical in ultrasound image segmentation for several reasons. First, precise delineation of organ and lesion boundaries directly impacts clinical diagnostic accuracy and treatment planning, as even minor boundary inaccuracies can lead to misdiagnosis or inappropriate therapeutic interventions. Second, the inherently low signal-to-noise ratio and tissue-dependent acoustic impedance variations in ultrasound imaging make boundary regions the most vulnerable areas for segmentation errors. Third, unlike natural images where edges are typically defined by sharp intensity transitions, ultrasound boundaries often manifest as gradual intensity changes or discontinuous edge fragments, requiring specialized mechanisms to capture these subtle boundary cues. Existing PEFT methods (e.g., LoRA \cite{hu2021lora}), while effective for general adaptation tasks, are not specifically designed to address the unique challenges of ultrasound imaging, particularly the critical need for enhanced boundary perception and contextual understanding in this domain. To address these limitations, we design a Context-Edge Hybrid Adapter (CH-Adapter) using the PEFT method to reduce trainable parameters while compensating for cross-domain differences between natural and ultrasound images, retaining essential pre-trained knowledge while updating only a minimal set of parameters. This design philosophy directly targets SAM2's core limitation in that its pre-trained representations, optimized for natural images, lack the domain-specific knowledge required to handle ultrasound's unique imaging physics and boundary ambiguity. Different CH-Adapters are inserted into the Hiera blocks in the image encoder at corresponding levels, thereby enhancing segmentation capability by integrating local and global attention mechanisms.  Our proposed CH-Adapter consists of context-aware and edge-aware components, focusing on extracting domain-specific information and injecting it into the backbone network. The context-aware component injects task-related contextual information, addressing the performance limitations caused by domain differences. The edge-aware component enhances object boundaries in feature maps, thus alleviating the segmentation challenges caused by the blurred boundaries characteristic of ultrasound images.

As illustrated in Figure \ref{fig:2}(b), for the context-aware component, the information $F_{l}$ obtained by the $l$-th CH-Adapter passes through the down projection layer, $\rm GELU$ \cite{hendrycks2016gaussian} layer and the up projection layer in sequence, and finally obtains the prompt $P^{l}$:
\begin{equation}
    P^{l}={\rm Linear}^l_{up}\left({\rm GELU}({\rm Linear}_{down}^{l}(F_{l}))\right),
\end{equation}
where ${\rm Linear}_{down}^{l}$ denotes the linear layers that map features from the input dimension to the lower dimension and are used to generate specific prompts for $l$-th CH-Adapter, $l=1,2,3$. ${\rm GELU}$ stands for the GELU activation function, ${\rm Linear}^l_{up}$ is a linear layer for up projection to raise the dimension, and $P^{l}$ is the output prompt attached to the Hiera Block \cite{ravi2024sam}. The context-aware component processes the hybrid image embedding through a structured pathway. The input features undergo dimensional transformation via the down projection layer, which effectively reduces computational complexity while preserving essential contextual information. This compressed representation is then enhanced through the GELU activation function, which introduces non-linearity crucial for complex feature learning, before being restored to the original dimensionality through the up projection layer.

The edge-aware component, as depicted in the lower portion of Figure \ref{fig:2}(b), is designed to enhance the model's ability to extract boundary features, thereby sharpening the fuzzy edges of the region of interest in ultrasound images. To mitigate computational demands, the dimensionality of the patch embedding is reduced through a patch mixer, effectively lowering the feature space dimensionality while retaining essential information. Following this, the dimensionality-reduced features $F_{l}$ are split to $C$ feature maps as  $F_{l}=\{F_l^i\}_{i=1}^{C}$. Then, $F_l^i$ are subjected to $3\times3$ Sobel filtering \cite{kanopoulos1988design} in four directions (horizontal, vertical, and diagonal directions) to augment the edge details corresponding to each direction. The resulting edge-enhanced feature maps are concatenated and subsequently passed through a $1\times1$ channel mixer to increase the feature dimensions, facilitating further processing and analysis in the subsequent stages of the model. The process can be expressed as:

\begin{equation}
    F_{l} \leftarrow\mathcal{M}\bigg(\{\sum_{k=1}^n\gamma(S_k,F_l^i)\}_{i=1}^{C}\bigg),
\end{equation}

where $C$ is the number of channels, $i=1,2,...,C$, $k$ represents the sequence number of the Sobel filters, $S_k$ denotes the $k$-th Sobel filter, $\gamma$ is the convolutional operator, $n=4$ corresponding to the four Sobel filters in the edge-aware component, and $\mathcal{M}$ refers to a $1\times1$ channel mixer that we use to adjust the dimension of the features.

Our CH-Adapter consists of the above two components, making an element-wise addition as:
\begin{equation}
    H_{l}=P^l+F_{l},
\end{equation}
where $H_{l}$ denotes the feature embedding output by the $l$-th adapter, the context-aware component prompt $P^l$ provides a rich source of contextual knowledge related to the task, helping the model to adapt to different domains and tasks, and the edge-aware component's feature $F_{l}$ enhances the boundaries of objects in the feature maps, making it easier for the model to distinguish between different regions, especially in ultrasound images where boundaries can be blurred. By adding these two components together, $H_{l}$ is a crucial part in our architecture and becomes a comprehensive representation that both contains the necessary contextual information and highlights the important edge details.

\subsection{Deep-Supervised Knowledge Distillation}
The clinical application of ultrasound image segmentation requires the model to be not only accurate and generalizable but also lightweight and suitable for clinical deployment. To this end, we design a DSKD method specifically for ultrasound image segmentation. Unlike standard knowledge distillation that compresses general models, our DSKD is designed to distill ultrasound-specific knowledge from the domain-adapted SAM2 into a lightweight encoder suitable for point-of-care devices with severe computational constraints, while preserving the fine-grained boundary perception critical for ultrasound imaging. As presented in Figure \ref{fig:2}(c), we distill Hiera block ($D_2$), CH-Adapter ($D_3$), and the integration of Hiera block and CH-Adapter ($D_1$). The integration operation in our DSKD framework is defined as element-wise addition. The formula represents the loss function, which achieves knowledge transfer by minimizing the output difference between the teacher model and the student model:
\begin{equation}
\begin{split}
    &L_{{\rm DSKD}}=\sum_{l=1}^n\bigg(\left\|\mathcal{F}_{D_1}^{l}(\boldsymbol{x}_l)-f_{D_1}^{l}(\boldsymbol{x}_l)\right\|_2^2+ \\
&\left\|\mathcal{F}_{D_2}^{l}(\boldsymbol{x}_l)-f_{D_2}^{l}(\boldsymbol{x}_l)\right\|_2^2
+\left\|\mathcal{F}_{D_3}^{l}(\boldsymbol{x}_l)-f_{D_3}^{l}\boldsymbol{x}_l)\right\|_2^2\bigg),
\end{split}
\label{eq:5}
\end{equation}
where $\mathcal{F}_{D_1}^{l}$, $\mathcal{F}_{D_2}^{l}$ and $\mathcal{F}_{D_3}^{l}$ represent the output of the teacher model at different levels, $f_{D_1}^{l}$, $f_{D_2}^{l}$ and $f_{D_3}^{l}$ represent the output of the student model at the corresponding levels, $\boldsymbol{x}_l$ denotes the input of three hierarchical stages and $l=1,2,3$. The CH-Adapter enables the model to adapt to specific application scenarios more accurately. Through the Hiera block, the model can obtain richer feature information. Meanwhile, the integration process endows the model with stronger feature representation capabilities, which enables it to better adapt to downstream tasks. Therefore, we distill all three levels to enhance the model's performance in ultrasound segmentation.

The knowledge distillation architecture, as shown in Figure \ref{fig:2}(c), establishes a comprehensive teacher-student learning paradigm where the teacher model guides the student model through multi-level feature alignment. The distillation process operates simultaneously at three hierarchical levels with different spatial resolutions, ensuring that knowledge transfer occurs across multiple scales of feature representation. The neck layer is implemented as a two-layer CNN structure, consisting of a $3\times3$ convolutional layer followed by a $1\times1$ convolutional layer. It serves as a dimensional adaptation mechanism between the teacher and student models, facilitating effective knowledge transfer by feature distillation.

\subsection{Model Architecture and Training Pipeline}

To construct our UniUltra framework, we first adopt Hiera ViT \cite{ryali2023hiera} as the backbone image encoder for universal ultrasound segmentation, ensuring consistent feature learning across diverse ultrasound imaging modalities. As shown in Figure \ref{fig:2}(a), we load SAM2-L \cite{ravi2024sam} to initialize corresponding parameters. Following the official SAM2 implementation, we discard the last stage of the Hiera encoder and retain only the first three stages, which ensures full compatibility with SAM2's pre-trained weights while reducing computational complexity and model size.
The image is divided into patches and mapped into patch embedding vectors, where one branch is processed by the patch mixer for dimensionality reduction while the other enters the Hiera Block. We freeze the Hiera blocks (indicated by snowflake symbols) to preserve pre-trained knowledge, while inserting our novel CH-Adapters between Hiera blocks (marked with flame symbols) to achieve parameter-efficient fine-tuning in ultrasound domains. The CH-Adapters consist of context-aware and edge-aware components that enable task-specific adaptation without compromising the foundational capabilities of the backbone network. To optimize training efficiency, we compress the embedding dimension from 768 to 192 within the CH-Adapter, as the limited scale of medical data does not require excessively high dimensions. The image encoder is followed by the original SAM2 decoding workflow with memory attention mechanism and bounding box prompts to provide explicit spatial guidance for precise segmentation. The training of UniUltra consists of two sequential phases: (1) parameter-efficient fine-tuning phase of UniUltra-L, where only CH-Adapters are trained while keeping Hiera blocks frozen, and (2) knowledge distillation phase, where we introduce our novel DSKD technique to independently distill each layer in the image encoder at each stage. To build our student model, we introduce the UniUltra-mini architecture that follows Hiera and reduces the embedding dimension from 768 to 128 to minimize computational cost and the learning cost of hard-sample representations during distillation. All images used for training serve as samples for the distillation process, ensuring consistency between distillation input and fine-tuning input. The unified pipeline guarantees that the compressed model inherits domain-specific knowledge while achieving a significant parameter reduction compared to the original SAM2. Overall, this collaborative design not only avoids redundant computation but also enables effective knowledge transfer from large-scale natural image pre-training to specialized ultrasound segmentation, leading to robust performance with minimal computational overhead.

\section{Experiments}
In this section, we first introduce six public ultrasound image datasets, with four used for internal validation and two for external validation. Secondly, the details on the implementation are provided, including the experimental environment, computational resources, and training configurations. Then, we describe the metrics used to evaluate model performance. Next, we present the main results, including internal validation, external validation, computational costs, and qualitative results. Finally, we conduct three groups of ablation studies to demonstrate the importance for each proposed component.

\begin{table*}[!t]
    \centering
    \small
    \setlength\tabcolsep{4.5pt}
    \caption{Comparison of \textit{internal validation} with state-of-the-arts on ultrasound segmentation. FT: fine-tuned.}
    {\scalebox{1}{
    \begin{tabular}{l|c|ccc|ccc|ccc|ccc}
    \hline
    \multirow{2}{*}{Methods} & \multirow{2.5}{*}{FT} & \multicolumn{3}{c|}{BUSI} & \multicolumn{3}{c|}{DDTI} & \multicolumn{3}{c|}{TN3K} & \multicolumn{3}{c}{UDIAT}\\
    \cline{3-14}
    & & Dice & mIoU & HD & Dice & mIoU & HD & Dice & mIoU & HD & Dice & mIoU & HD\\
    \hline
    UNet \cite{ronneberger2015u} MICCAI'15 & \multirow{6}{*}{\ding{51}} & 60.29 & 47.69 & 352.21 & 65.06 & 49.97 & 280.81 & 61.60 & 48.32 & 370.25 & 56.55 & 44.85 & 309.08\\
    nnU-Net \cite{isensee2021nnu} NM'21 &  & 85.76 & 78.03 & 92.08 & 83.98 & 74.69 & 104.33 & 83.20 & 74.53 & 102.00 & 82.34 & 75.09 & 96.41\\
    AAU-Net \cite{chen2022aau} TMI'22 &  & 80.13 & 71.68 & 129.75 & 75.98 & 64.80 & 131.23 & 78.05 & 68.01 & 122.72 & 73.92 & 63.67 & 144.64\\
    TransUNet \cite{chen2024transunet} MedIA'24 &  & 73.62 & 62.15 & 131.97 & 72.98 & 60.63 & 102.60 & 74.61 & 63.53 & 108.94 & 70.84 & 59.09 & 107.80\\
    EMCAD \cite{rahman2024emcad} CVPR'24 &  & 79.35 & 70.08 & 133.36 & 76.46 & 66.18 & 137.69 & 79.20 & 69.67 & 125.22 & 80.53 & 71.22 & 133.11\\
    MADGNet \cite{nam2024modality} CVPR'24 &  & 80.00 & 70.58 & 157.17 & 75.63 & 64.40 & 185.00 & 77.39 & 67.37 & 152.71 & 71.86 & 61.12 & 176.02\\
    \hline
    SAM \cite{kirillov2023segment} ICCV'23 & \multirow{4}{*}{\ding{55}} & 87.45 & 78.29 & 80.77 & 85.58 & 75.15 & 156.77 & 79.92 & 69.98 & 104.68 & 89.45 & 81.54 & 69.78\\
    MedSAM \cite{ma2024segment} NC'24 &  & 93.11 & 87.34 & 54.52 & 90.56 & 82.98 & 56.50 & 90.29 & 87.27 & 45.48 & 90.07 & 83.12 & 41.54\\
    SAMMI \cite{huang2024segment} MedIA'24 &  & 91.96 & 85.46 & 61.22 & 91.70 & 84.81 & 81.61 & 89.24 & 82.06 & 70.83 & 92.13 & 85.73 & 49.97\\
    SAM2 \cite{ravi2024sam} ICLR'25 &  & 89.63 & 81.80 & 59.65 & 90.50 & 83.09 & 72.02 & 86.19 & 78.08 & 75.59 & 93.05 & 87.12 & 44.22\\
    \hline
    SAMed \cite{zhang2023customized} arXiv'23  & \multirow{5}{*}{\ding{51}} & 92.67 & 86.64 & 47.49 & 93.16 & 87.32 & 51.60 & 93.02 & 87.16 & 44.75 & 92.69 & 86.64 & 46.49\\
    SAM-CL \cite{zhong2024convolution} ICLR'24 &  & 92.82 & 86.79 & 58.11 & 93.48 & 87.84 & 51.14 & 93.61 & 88.16 & 38.57 & 92.64 & 86.62 & 45.02 \\
    SAMUS \cite{lin2024beyond} MICCAI'24 &  & 92.86 & 86.88 & 55.61 & 93.46 & 87.83 & 49.69 & 93.56 & 88.08 & 37.93 & 93.22 & 87.49 & 44.55 \\
    Med-SA \cite{wu2023medical} MedIA'25 &  & 89.84 & 82.03 & 60.24 & 91.68 & 84.81 & 58.82 & 91.38 & 84.41 & 49.16 & 89.03 & 80.59 & 63.27\\
    UniUltra-L (Ours) &  & \textbf{93.81} & \textbf{88.48} & \textbf{42.67} & \textbf{94.20} &\textbf{89.13} & \textbf{43.93} & \textbf{94.50} & \textbf{89.70} & \textbf{34.27} & \textbf{94.15} & \textbf{89.06} & \textbf{36.98}\\
    \hline
    MobileSAM \cite{zhang2023faster} arXiv'23 & \multirow{5}{*}{\ding{51}}  & 92.68 & 86.61 & 52.92 & 93.44 & 87.79 & 51.50 & 93.67 & 88.28 & 43.02 & 92.96 & 87.03 & 45.80 \\
    EfficientSAM \cite{xiong2024efficientsam} CVPR'24 & & 89.47 & 81.91 & 77.82 & 91.30 & 84.16 & 79.86 & 88.64 & 81.01 & 81.94 & 90.55 & 83.03 & 57.62 \\
    EdgeSAM \cite{zhang2024efficientvit} CVPRW'24 & & 92.73 & 86.67 & 54.03 & 92.89 & 86.84 & 51.86 & 92.53 & 86.45 & 46.59 & 93.28 & 87.61 & 45.13 \\
    RepViT-SAM \cite{wang2023repvit} CVPR'24 & & 92.42 & 86.14 & 56.10 & 93.35 & 87.62 & 59.86 & 92.47 & 86.34 & 52.63 & 93.31 & 87.63 & 41.39 \\
    TinySAM \cite{shu2023tinysam} AAAI'25 &  & 93.12 & 87.27 & 53.53 & 92.63 & 86.44 & 59.54 & 92.97 & 87.09 & 46.96 & 92.18 & 85.68 & 45.86 \\
    UniUltra-mini (Ours) &  & \textbf{93.21} & \textbf{87.44} & \textbf{49.13} & \textbf{94.04} & \textbf{88.84} & \textbf{45.12} & \textbf{93.99} & \textbf{88.79} & \textbf{35.47} & \textbf{93.82} & \textbf{88.46} & \textbf{40.41}\\
    \hline
    \end{tabular}}}
    \label{tab:1}
    \vspace{-1.0em}
\end{table*}

\subsection{Datasets}
To validate the effectiveness of the proposed UniUltra, we collect four public datasets named BUSI \cite{al2020dataset}, DDTI \cite{pedraza2015open}, TN3K \cite{gong2023thyroid} and UDIAT \cite{yap2017automated} for internal validation to assess the method performance in different tasks. We adopt a common 8:1:1 train-val-test with a random split, where the data is initially randomly partitioned and then fixed throughout all experiments to ensure reproducibility. The validation set (10\%) is used for hyperparameter tuning and model selection during training, while the test set (10\%) is strictly held out for final performance evaluation, with no overlap between splits to prevent data leakage. In addition, we conduct external validation on two datasets FUGC \footnote{https://www.codabench.org/competitions/4781/\label{fugc}} and JNU-IFM \cite{lu2022jnu} for evaluating generalization capabilities.

 \noindent\textbf{BUSI} \cite{al2020dataset} is a breast ultrasound segmentation dataset collected from 600 female patients, and contains 780 images of $500\times500$ resolutions. It is divided into three categories: normal, benign and malignant. In our experiments, we include images from both benign and malignant classes, with a total of 647 images across the two categories.

 \noindent\textbf{DDTI} \cite{pedraza2015open} is a thyroid nodule ultrasound segmentation dataset released by the National University of Colombia, comprising 637 images collected from a single device. The image size varies from $245\times360$ to $560\times360$.

 \noindent\textbf{TN3K} \cite{gong2023thyroid} is a thyroid nodule ultrasound segmentation dataset from 2421 patients, and contains 3493 images obtained from multiple devices and perspectives. The image size ranges from $216\times 217$ to $1463\times771$.

 \noindent\textbf{UDIAT} \cite{yap2017automated} is a breast ultrasound segmentation dataset consisting of benign and malignant cases. It contains 163 images, with dimensions ranging from $344\times233$ to $753\times617$.

 \noindent\textbf{FUGC} \textsuperscript{\ref{fugc}} is a cervical ultrasound segmentation dataset deriving from the fetal ultrasound grand challenge. It consists of 500 images (labeled 50 images) with a fixed resolution of $544\times 366$. In our experiments, we focus on the 50 images with pixel-level annotations. The images contained two classes of information. We manually divide them into two categories: Anterior lip as FUGC-AL and Posterior lip as FUGC-PL.

 \noindent\textbf{JNU-IFM} \cite{lu2022jnu} is a public symphysis-fetal head ultrasound segmentation dataset. It contains 6224 frame images of $1295\times1026$ resolution from 51 videos of 51 patients. We manually divide them into two classes: Head as JNU-IFM-Head and Symphysis as JNU-IFM-SP.

\begin{table*}[!t]
    \centering
    \small
    \setlength\tabcolsep{5.3pt}
    \caption{Comparison of \textit{external validation} with state-of-the-arts on ultrasound segmentation.}
    {\scalebox{1}{
    \begin{tabular}{l|ccc|ccc|ccc|ccc}
    \hline
    \multirow{2}{*}{Methods} & \multicolumn{3}{c|}{FUGC-AL} & \multicolumn{3}{c|}{FUGC-PL} & \multicolumn{3}{c|}{JNU-IFM-Head} & \multicolumn{3}{c}{JNU-IFM-SP}\\
    \cline{2-13}
    & Dice & mIoU & HD & Dice & mIoU & HD & Dice & mIoU & HD & Dice & mIoU & HD \\
    \hline
    UNet \cite{ronneberger2015u} MICCAI'15 & 12.04 & 6.53 & 570.27 & 13.62 & 7.69 & 490.94 & 44.48 & 29.91 & 502.79 & 2.04 & 1.04 & 587.35 \\
    nnU-Net \cite{isensee2021nnu} NM'21 & 14.65 & 8.46 & 311.28 & 5.96 & 3.94 & 476.13 & 36.22 & 27.97 & 276.68 & 26.87 & 17.26 & 349.63 \\
    AAU-Net \cite{chen2022aau} TMI'22 & 4.72 & 2.49 & 375.64 & 21.47 & 14.24 & 457.36 & 26.47 & 20.28 & 324.61 & 3.20 & 1.92 & 480.91 \\
    TransUNet \cite{chen2024transunet} MedIA'24 & 26.25 & 15.84 & 342.24 & 13.41 & 7.67 & 402.64 & 49.14 & 34.09 & 304.72 & 5.80 & 3.06 & 510.92 \\
    EMCAD \cite{rahman2024emcad} CVPR'24 & 27.31 & 17.41 & 340.50 & 18.60 & 12.41 & 386.83 & 34.63 & 24.74 & 286.59 & 13.64 & 7.87 & 411.94 \\
    MADGNet \cite{nam2024modality} CVPR'24 & 27.76 & 16.88 & 385.03 & 27.74 & 18.86 & 431.24 & 38.97 & 27.58 & 336.58 & 7.91 & 4.31 & 514.79 \\
    \hline
    SAM \cite{kirillov2023segment} ICCV'23 & 42.46 & 31.65 & 301.99 & 72.01 & 56.51 & 235.81 & 86.77 & 77.64 & 90.78 & 61.54 & 46.54 & 48.02 \\
    MedSAM \cite{ma2024segment} NC'24 & 73.51 & 58.93 & 241.19 & 81.10 & 68.89 & 139.08 & 93.57 & 88.02 & 43.11 & 83.29 & 74.62 & 41.84 \\
    SAMMI \cite{huang2024segment} MedIA'24 & 65.26 & 49.58 & 277.62 & 73.44 & 58.31 & 185.28 & 92.36 & 86.33 & 64.34 & 81.08 & 70.21 & 48.05 \\
    SAM2 \cite{ravi2024sam} ICLR'25 & 66.34 & 52.63 & 298.29 & 69.55 & 55.62 & 215.27 & 88.32 & 80.40 & 74.59 & 79.17 & 66.95 & 38.06 \\
    \hline
    SAMed \cite{zhang2023customized} arXiv'23& 67.93 & 51.84 & 256.82 & 77.40 & 63.45 & 152.03 & 94.83 & 90.23 & 36.31 & 87.96 & 79.74 & 24.07 \\
    SAM-CL \cite{zhong2024convolution} ICLR'24& 71.83 & 56.31 & 319.58 & 74.11 & 59.14 & 180.17 & 93.08 & 87.21 & 47.87 & 85.54 & 76.06 & 30.23 \\
    SAMUS \cite{lin2024beyond} MICCAI'24 & 67.52 & 51.24 & 269.84 & 78.48 & 64.95 & 144.73 & 93.86 & 88.55 & 42.39 & 87.62 & 79.42 & 24.17 \\
    Med-SA \cite{wu2023medical} MedIA'25 & 66.73 & 50.21 & 276.83 & 71.56 & 55.94 & 167.30 & 95.99 & 92.33 & 28.12 & 87.27 & 78.77 & 25.05 \\
    UniUltra-L (Ours) & \textbf{75.22} & \textbf{62.86} & \textbf{158.26} & \textbf{82.07} & \textbf{69.92} & \textbf{138.49} & \textbf{96.37} & \textbf{93.71} & \textbf{19.44} & \textbf{90.83} & \textbf{84.17} & \textbf{13.83}\\ 
    \hline
    MobileSAM \cite{zhang2023faster} arXiv'23 & 69.77 & 54.15 & 252.45 & 76.24 & 61.90 & 175.23 & 93.52 & 87.94 & 45.16 & 87.68 & 79.35 & 25.61 \\
    EfficientSAM \cite{xiong2024efficientsam} CVPR'24 & 60.13 & 43.47 & 285.35 & 69.77 & 53.81 & 226.66 & 93.37 & 87.74 & 85.78 & 69.42 & 54.88 & 123.53 \\
    EdgeSAM \cite{zhang2024efficientvit} CVPRW'24 & 73.07 & 58.17 & 248.26 & 76.09 & 61.71 & 146.94 & 91.95 & 85.22 & 53.93 & 86.67 & 77.44 & 22.92 \\
    RepViT-SAM \cite{wang2023repvit} CVPR'24 & 71.50 & 56.35 & 222.37 & 78.59 & 65.22 & 151.90 & 93.43 & 87.77 & 44.45 & 86.72 & 77.66 & 23.88 \\
    TinySAM \cite{shu2023tinysam} AAAI'25 & 67.30 & 51.18 & 292.02 & 75.57 & 61.05 & 196.13 & 93.25 & 87.48 & 47.28 & 83.59 & 73.29 & 72.69 \\

    UniUltra-mini (Ours) & \textbf{74.33} & \textbf{60.41} & \textbf{170.45} & \textbf{79.43} &\textbf{66.27} & \textbf{140.27} & \textbf{94.95} & \textbf{91.45} & \textbf{23.32} & \textbf{89.89} & \textbf{82.25} & \textbf{15.72}\\
    \hline
    \end{tabular}}}
    \label{tab:2}
    \vspace{-1.0em}
\end{table*}

\subsection{Implementation Details}
We conduct our experiments on four parallel NVIDIA RTX A5000 GPUs (24GB), utilizing PyTorch 2.4.1, Python 3.10, and CUDA 12.4. To guarantee experimental fairness and reproducibility, we maintain the same training settings and configurations throughout all models. We utilize Adam as the optimizer with a batch size of 4 and 200 epochs. For the learning rate, we initialize it at 0.0001 and employ an exponential decay strategy with a factor of 0.98 for adjustment. All images are resized to 1024 $\times$ 1024 during both training and testing phases. For bounding box prompt generation, we follow the methods employed in MedSAM \cite{ma2024segment} and SAMMI \cite{huang2024segment}. The bounding box is automatically extracted as the minimum enclosing rectangle of the ground truth mask with a random perturbation of 0-20 pixels applied to the coordinates. For evaluating the upper bound performance of models, we select the ViT-H and Hiera-L as the image encoder for all SAM and SAM2 frameworks, respectively. The predicted segmentation mask is supervised by the weighted combination of focal loss and dice loss \cite{kirillov2023segment}. 

\subsection{Evaluation Metrics}
In our experiments, we use three metrics, Dice coefficient, mean Intersection over Union (mIoU), and Hausdorff distance (HD), to evaluate the performance of different models on ultrasound image segmentation through internal and external validation. The Dice coefficient serves as the primary metric for evaluating segmentation overlap, providing a measure of spatial agreement between predicted and ground truth masks. The mean Intersection over Union (mIoU) offers complementary information about region-based accuracy. The Hausdorff Distance (HD) specifically quantifies boundary precision, which is particularly critical in medical applications where precise delineation of anatomical structures is essential for clinical decision-making. In addition, we report the number of Trainable Parameters (TP), the number of parameters of the image encoder (Params), Floating Point of Operations (FLOPs), and Frames Per Second (FPS).

\subsection{Main Results}

We compare our UniUltra and SOTA methods on both internal and external validation settings.

\subsubsection{Comparison on Internal Validation}
As illustrated in Table \ref{tab:1}, experimental results reveal distinct performance characteristics across model categories. Traditional UNet-based architectures demonstrate substantial limitations, with UNet achieving Dice scores ranging from 56.55\% to 65.06\% and HD values exceeding 280 pixels across all datasets. nnU-Net emerges as the most robust traditional approach, achieving Dice scores between 82.34\% and 85.76\%, though HD performance remains suboptimal with values ranging from 92.08 to 104.33 pixels. SAM-based foundation models show better performance than traditional UNet variants, with MedSAM demonstrating substantial improvements over baseline methods. Compared with nnU-Net, Dice of MedSAM is 7.35\%, 6.58\%, 7.09\% and 7.73\% higher on four datasets, while achieving HD values between 41.54 and 56.50 pixels. Nevertheless, PEFT and lightweight SAMs outperform MedSAM on most datasets, with models such as SAMed, SAM-CL, and SAMUS achieving Dice scores consistently above 92\% and HD values below 60 pixels.

The proposed UniUltra surpasses all SOTA models across all evaluation metrics. UniUltra-L achieves exceptional performance with Dice scores ranging from 93.81\% to 94.50\% and HD values between 34.27 and 43.93 pixels (with a P-value $<$ 0.005). Compared to the best-performing competitors on each dataset (TinySAM on BUSI, SAM-CL on DDTI, MobileSAM on TN3K, and RepViT-SAM on UDIAT), UniUltra-L shows Dice improvements of 0.69\% to 0.84\% and HD reductions of 9.65\% to 11.59\% (relative to SAMed on BUSI, SAMUS on DDTI and TN3K, and RepViT-SAM on UDIAT). Compared to MedSAM across all datasets, UniUltra-L demonstrates consistent improvements with Dice gains ranging from 0.70\% to 4.21\% and HD reductions from 10.98\% to 24.65\% (with a P-value $<$ 0.005). On BUSI dataset, UniUltra-L achieves 93.81\% Dice, which is 0.75\% higher than MedSAM's 93.11\%, while reducing HD from 54.52 to 42.67 pixels (21.74\% improvement). UniUltra-mini maintains competitive performance with Dice scores between 93.21\% and 94.04\% while achieving HD values from 35.47 to 49.13 pixels, consistently outperforming most methods including MedSAM (with a P-value $<$ 0.001). These comprehensive comparisons demonstrate UniUltra's superior segmentation accuracy and boundary detection capabilities.

\begin{figure*}[!t]
  \centering
  \includegraphics[width=0.95\linewidth]{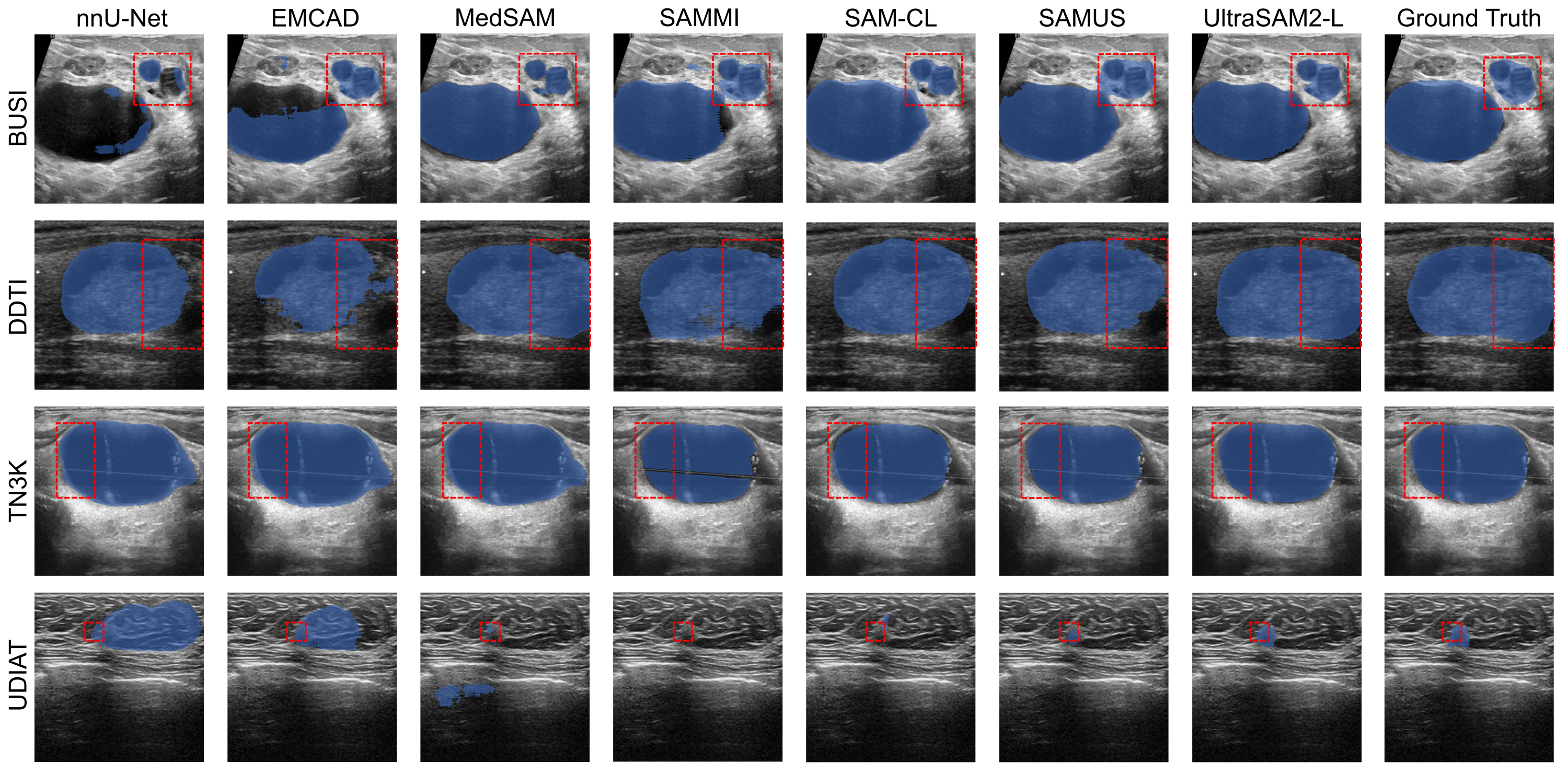}
  \caption{Visualization of universal ultrasound segmentation on interval validation. Our UniUltra exhibits the best results, recognizing more accurate lesion regions with delineating precise boundaries while having fewer false positives.}
  \label{fig:m3}

\end{figure*}

\begin{table}[!t]
    \centering
    \small
    \setlength\tabcolsep{4pt}
    \caption{Computational Efficiency Comparison of SAM-based Models}
    \begin{tabular}{l|cccc}
    \hline
    Methods & Params (M) & FLOPs (G) & FPS & Dice \\
    \hline
    SAM \cite{kirillov2023segment} & 633.03 & 2739.76 & 2.15 & 75.65 \\
    MedSAM \cite{ma2024segment} & 86.67 & 369.01 & 7.42 & 86.94 \\
    SAMMI \cite{huang2024segment} & 634.03 & 2736.61 & 1.07 & 84.64 \\
    SAM2 \cite{ravi2024sam} & 209.70 & 810.99 & 17.82 & 82.84 \\
    SAMed \cite{zhang2023customized} & 638.75 & 2739.76 & 2.27 & 87.46 \\
    SAM-CL \cite{zhong2024convolution} & 705.49 & 3148.25 & 1.43 & 87.14 \\
    SAMUS \cite{lin2024beyond} & 747.36 & 3574.21 & 1.11 & 87.57 \\
    Med-SA \cite{wu2023medical} & 690.62 & 3086.30 & 1.94 & 85.44 \\
    MobileSAM \cite{zhang2023faster} & 6.07 & 36.74 & 23.93 & 87.49 \\
    EfficientSAM \cite{xiong2024efficientsam} & 6.16 & 25.04 & 27.14 & 81.58 \\
    EdgeSAM \cite{zhang2024efficientvit} & 5.47 & 22.10 & 30.57 & 87.40 \\
    RepViT-SAM \cite{wang2023repvit} & 5.12 & 18.61 & 28.12 & 87.72 \\
    TinySAM \cite{shu2023tinysam} & 6.07 & 36.74 & 20.07 & 86.33 \\
    \hline
    UniUltra-L (Ours) & 230.59 & 847.47 & 16.21 & \textbf{90.14} \\
    UniUltra-mini (Ours) & \textbf{0.86} & \textbf{7.38} & \textbf{45.09} & 89.21 \\
    \hline
    \end{tabular}
    \label{tab:6}
\end{table}

\subsubsection{Comparison on External Validation}
To evaluate the generalization capability of our UniUltra, we conduct external validation on two additional public ultrasound image datasets. The optimal weights trained on internal datasets are directly applied to external datasets without any fine-tuning, providing a rigorous assessment of zero-shot generalization performance. As shown in Table \ref{tab:2}, FUGC-AL and FUGC-PL represent the anterior lip and posterior lip categories of the FUGC dataset, while JNU-IFM-Head and JNU-IFM-SP correspond to the fetal head and symphysis pubis categories in the JNU-IFM dataset. The external validation results demonstrate that UNet-based models exhibit the poorest performance, rendering them unsuitable for practical deployment, with UNet achieving Dice scores ranging from 2.04\% to 44.48\% and HD values exceeding 490 pixels across all categories. In contrast, SAM-based models, PEFT SAMs, and lightweight SAMs demonstrate substantially better performance than UNet-based approaches, with the original SAM achieving Dice scores between 42.46\% and 86.77\%. Among these models, MedSAM exhibits superior performance due to extensive pre-training on medical imaging datasets, achieving Dice scores ranging from 73.51\% to 93.57\% across four external validation categories.

Nevertheless, our proposed UniUltra demonstrates exceptional performance on both external datasets irrespective of model size, surpassing all SOTA models across all evaluation metrics. UniUltra-L achieves outstanding performance with Dice scores ranging from 75.22\% to 96.37\% and HD values between 13.83 and 158.26 pixels (with a P-value $<$ 0.001). Specifically, compared to MedSAM, the best-performing baseline model, UniUltra-L achieves an average improvement of 3.26\% in Dice score and 39.17\% reduction in HD across both datasets. Relative to the original SAM, UniUltra-L demonstrates significant Dice improvements of 32.76\%, 10.06\%, 9.60\%, and 29.29\% on FUGC-AL, FUGC-PL, JNU-IFM-Head, and JNU-IFM-SP categories, respectively. UniUltra-mini maintains remarkable generalization capabilities despite significant parameter reduction, achieving Dice scores between 74.33\% and 94.95\% and HD values from 15.72 to 170.45 pixels (with a P-value $<$ 0.001). Compared to SAM2, UniUltra-mini achieves average enhancements of 8.81\% in Dice score and 51.27\% in HD performance across all categories. These comprehensive comparisons conclusively validate the robust generalization capability of our proposed model on previously unseen datasets.

\begin{table}[!t]
\centering
\setlength\tabcolsep{13pt}
\caption{Memory consumption comparison during inference.}
\label{tab:memory}
\adjustbox{width=0.48\textwidth,center}{\begin{tabular}{l|c}
\hline
Methods & GPU Memory Consumption (MB) \\
\hline
SAM \cite{kirillov2023segment} & 77,083 \\
SAM2 \cite{ravi2024sam} & 33,821 \\
UNet \cite{ronneberger2015u} & 29,511 \\
\hline
UniUltra-L (Ours) & \textbf{25,759} \\
UniUltra-mini (Ours) & \textbf{9,881} \\
\hline
\end{tabular}}
\end{table}

\begin{table}[!t]
    \centering
    \small
    \setlength\tabcolsep{3pt}
    \caption{Ablation study of CH-Adapter with internal validation. TP: trainable parameters.}
    {\scalebox{1}{
    \begin{tabular}{l|ccc|cc}
    \hline
    Methods & Dice & mIoU & HD & TP (M) & Ratio (\%)\\
    \hline
    Decoder Only & 91.05 & 84.03 & 56.47 & 11.73 & 5.23\\
    \hline
    LoRA \cite{hu2021lora} & 92.94 & 87.18 & 49.52 & 12.15 & 5.40\\
    VPT \cite{jia2022visual} & 92.71 & 86.95 & 51.38 & 13.57 & 5.74\\
    E$^2$VPT \cite{han20232vpt} & 93.24 & 87.55 & 49.02 & 15.87 & 6.64\\
    Adapter \cite{chen2024sam2} & 93.38 & 87.69 & 45.84 & 20.66 & 8.85\\
    Adapter + LoRA \cite{chen2024ma} & 93.51 & 87.88 & 43.56 & 21.08 & 9.02\\
    Conv-LoRA \cite{zhong2024convolution} & 93.47 & 87.82 & 44.19 & 14.23 & 6.01\\
    \hline
    CH-Adapter (Ours) & 94.17 & 89.09 & 39.46 & 20.80 & 8.91\\
    \hline
    \end{tabular}}}
    \label{tab:3}
\end{table}

\subsubsection{Comparison on Computational Costs}
The computational efficiency comparison presented in Table \ref{tab:6} demonstrates the performance trade-offs among various SAM-based models across four key metrics: Params of the image encoder, FLOPs, FPS, and average Dice. The original SAM model exhibits substantial computational overhead with 633.03M parameters (image encoder) and 2739.76G FLOPs, achieving only 2.15 FPS while maintaining a Dice score of 75.65\%. Medical-domain adaptations such as MedSAM, SAMed, and SAMUS show improved segmentation accuracy with Dice scores ranging from 86.94\% to 87.57\%, but most retain the computational burden of the original architecture, with parameter counts ranging from 86.67M to 747.36M and FLOPs ranging from 369.01G to 3574.21G. The efficiency-oriented variants present a contrasting profile, with MobileSAM, EfficientSAM, EdgeSAM, RepViT-SAM, and TinySAM achieving dramatic reductions in computational requirements. These models maintain parameter counts below 6.2M and FLOPs under 37G, resulting in substantially higher inference speeds ranging from 20.07 to 30.57 FPS. Notably, EdgeSAM achieves the highest inference speed at 30.57 FPS with only 5.47M parameters, while RepViT-SAM demonstrates the most efficient FLOP utilization at 18.61G. Despite their computational efficiency, these lightweight models maintain competitive segmentation performance, with average Dice scores between 81.58\% and 87.72\% across eight ultrasound segmentation categories of six different datasets.

Our proposed UniUltra variants establish a new paradigm in the efficiency-accuracy trade-off spectrum. UniUltra-L achieves the highest segmentation accuracy among all methods with a Dice score of 90.14\%, while maintaining reasonable computational efficiency with 230.59M parameters and 847.47G FLOPs. More remarkably, UniUltra-mini represents an extreme efficiency optimization, utilizing only 0.86M parameters and 7.38G FLOPs while achieving 45.09 FPS and maintaining a competitive Dice score of 89.21\%. This represents a 99.86\% reduction in parameters and 99.73\% reduction in FLOPs compared to the original SAM, while simultaneously improving segmentation accuracy by 13.56 percentage points and increasing inference speed by 20.97$\times$. The superior performance of both UniUltra variants validates our architectural innovations and optimization strategies by demonstrating that substantial efficiency gains can be achieved while simultaneously improving segmentation quality.

As shown in Table~\ref{tab:memory}, our UniUltra-L model achieves competitive memory efficiency with only 25,759 MB. More importantly, our lightweight variant UniUltra-mini requires only 9,881 MB of memory, representing an 87.2\% reduction compared to SAM and a 70.8\% reduction compared to SAM2. This remarkable memory efficiency makes UniUltra-mini particularly suitable for deployment in resource-constrained clinical environments, where GPU memory is often limited.

\begin{table}[!t]
    \centering
    \small
    \setlength\tabcolsep{5.5pt}
    \caption{Analysis of edge-aware efficiency with internal validation.}
    {\scalebox{1}{
    \begin{tabular}{cccc|ccc}
    \hline
    Horz. & Vert. & Diag. (R) & Diag. (L) & Dice & mIoU & HD\\
    \hline
     &  &  &  & 93.38 & 87.69 & 45.84\\ 
    \checkmark &  &  &  & 93.71 & 88.19 & 42.91\\ 
    \checkmark & \checkmark &  &  & 93.84 & 88.35 & 41.12\\
    \checkmark & \checkmark & \checkmark & & 93.98 & 88.59 & 40.47\\
    \checkmark & \checkmark & \checkmark & \checkmark & 94.17 & 89.09 & 39.46\\
    \hline
    \end{tabular}}}
    \label{tab:4}
\end{table}

\begin{table}[!t]
    \centering
    \small
    \setlength\tabcolsep{6.8pt}
    \caption{Ablation study of DSKD with external validation.}
    {\scalebox{1}{
    \begin{tabular}{ccc|ccc}
    \hline
    Integration & Hiera & CH-Adapter & Dice & mIoU & HD\\
    \hline
    \checkmark &  &  &   92.35 & 86.18 & 58.87\\ 
    \checkmark & \checkmark &  &   93.08 & 87.06 & 50.33\\
    \checkmark &  & \checkmark &   93.27 & 87.61 & 48.64\\
    \checkmark & \checkmark & \checkmark  & 93.77 & 88.38 & 42.53\\
    \hline
    \end{tabular}}}
    \label{tab:5}

\end{table}

\begin{table*}[!t]
\centering
\setlength\tabcolsep{8pt}
\caption{Comparison of Different Edge Operators on Internal Validation Datasets.}
\label{tab:edge_operator_comparison}
\renewcommand{\arraystretch}{1.2}
\adjustbox{width=1\textwidth,center}{\begin{tabular}{l|ccc|ccc|ccc|ccc}
\hline
\multirow{2}{*}{Edge Operator} & \multicolumn{3}{c|}{BUSI} & \multicolumn{3}{c|}{DDTI} & \multicolumn{3}{c|}{TN3K} & \multicolumn{3}{c}{UDIAT} \\
\cline{2-4} \cline{5-7} \cline{8-10} \cline{11-13}
 & Dice & mIoU & HD & Dice & mIoU & HD & Dice & mIoU & HD & Dice & mIoU & HD \\
\hline
Laplace & 92.61 & 86.39 & 47.53 & 94.08 & 88.92 & 42.68 & 94.40 & 89.52 & 38.48 & 92.97 & 87.02 & 31.18 \\
Canny & 92.78 & 86.71 & 47.53 & 94.12 & 89.04 & 44.22 & \textbf{94.66} & \textbf{89.97} & 37.41 & 92.94 & 86.93 & \textbf{30.36} \\
Sobel &  \textbf{93.81} & \textbf{88.48} & \textbf{42.67} & \textbf{94.20} &\textbf{89.13} & \textbf{43.93} & 94.50 & 89.70 & \textbf{34.27} & \textbf{94.15} & \textbf{89.06} & 36.98\\
\hline
\end{tabular}}
\end{table*}

\begin{table*}[!t]
\centering
\setlength\tabcolsep{8pt}
\caption{Comparison of Different Prompt Methods on Internal Validation Datasets.}
\label{tab:prompt_comparison}
\adjustbox{width=1\textwidth,center}{\begin{tabular}{l|ccc|ccc|ccc|ccc}
\hline
\multirow{2}{*}{Prompt Method} & \multicolumn{3}{c|}{BUSI} & \multicolumn{3}{c|}{DDTI} & \multicolumn{3}{c|}{TN3K} & \multicolumn{3}{c}{UDIAT} \\
\cline{2-4} \cline{5-7} \cline{8-10} \cline{11-13}
 & Dice & mIoU & HD & Dice & mIoU & HD & Dice & mIoU & HD & Dice & mIoU & HD \\
\hline
No prompt & 79.84 & 71.43 & 118.74 & 83.49 & 73.80 & 97.30 & 82.79 & 74.20 & 100.70 & 82.88 & 75.28 & 62.10 \\
1 point & 89.38 & 81.88 & 64.59 & 90.14 & 82.68 & 59.26 & 89.09 & 81.68 & 65.45 & 91.49 & 84.51 & 44.08 \\
3 points & 88.79 & 81.03 & 58.81 & 88.99 & 82.09 & 62.03 & 89.68 & 82.55 & 51.43 & 92.42 & 86.07 & 43.02 \\
1 bbox & 93.81 & 88.48 & 42.67 & 94.20 & 89.13 & 43.93 & 94.50 & 89.70 & 34.27 & 94.15 & 89.06 & 36.98\\
1 point + 1 bbox & \textbf{94.23} & \textbf{89.01} & \textbf{39.85} & \textbf{94.49} & \textbf{89.63} & \textbf{38.77} & \textbf{94.72} & \textbf{90.08} & \textbf{32.15} & \textbf{94.28} & \textbf{90.19} & \textbf{35.26} \\
\hline
\end{tabular}}
\end{table*}

\begin{table*}[!t]
\centering
\setlength\tabcolsep{8pt}
\caption{Ablation Study of DSKD with different numbers of distillation layers.}
\label{tab:level_distill}
\renewcommand{\arraystretch}{1.2}
\adjustbox{width=1\textwidth,center}{\begin{tabular}{c|ccc|ccc|ccc|ccc}
\hline
Distillation & \multicolumn{3}{c|}{BUSI} & \multicolumn{3}{c|}{DDTI} & \multicolumn{3}{c|}{TN3K} & \multicolumn{3}{c}{UDIAT} \\
\cline{2-4} \cline{5-7} \cline{8-10} \cline{11-13}
Layers & Dice & mIoU & HD & Dice & mIoU & HD & Dice & mIoU & HD & Dice & mIoU & HD \\
\hline
1 & 91.18 & 84.03 & 55.47 & 92.71 & 86.52 & 56.57 & 92.34 & 86.01 & 54.66 & 91.07 & 83.75 & 51.42 \\
2 & 92.45 & 86.16 & 52.27 & 93.40 & 87.71 & 48.78 & 93.23 & 87.51 & 43.44 & 92.94 & 86.94 & 46.41 \\
3 & \textbf{93.21} & \textbf{87.44} & \textbf{49.13} & \textbf{94.04} & \textbf{88.84} & \textbf{45.12} & \textbf{93.99} & \textbf{88.79} & \textbf{35.47} & \textbf{93.82} & \textbf{88.46} & \textbf{40.41}\\
\hline
\end{tabular}}
\end{table*}

\begin{figure}[!t]
  \centering
  \includegraphics[width=1\linewidth]{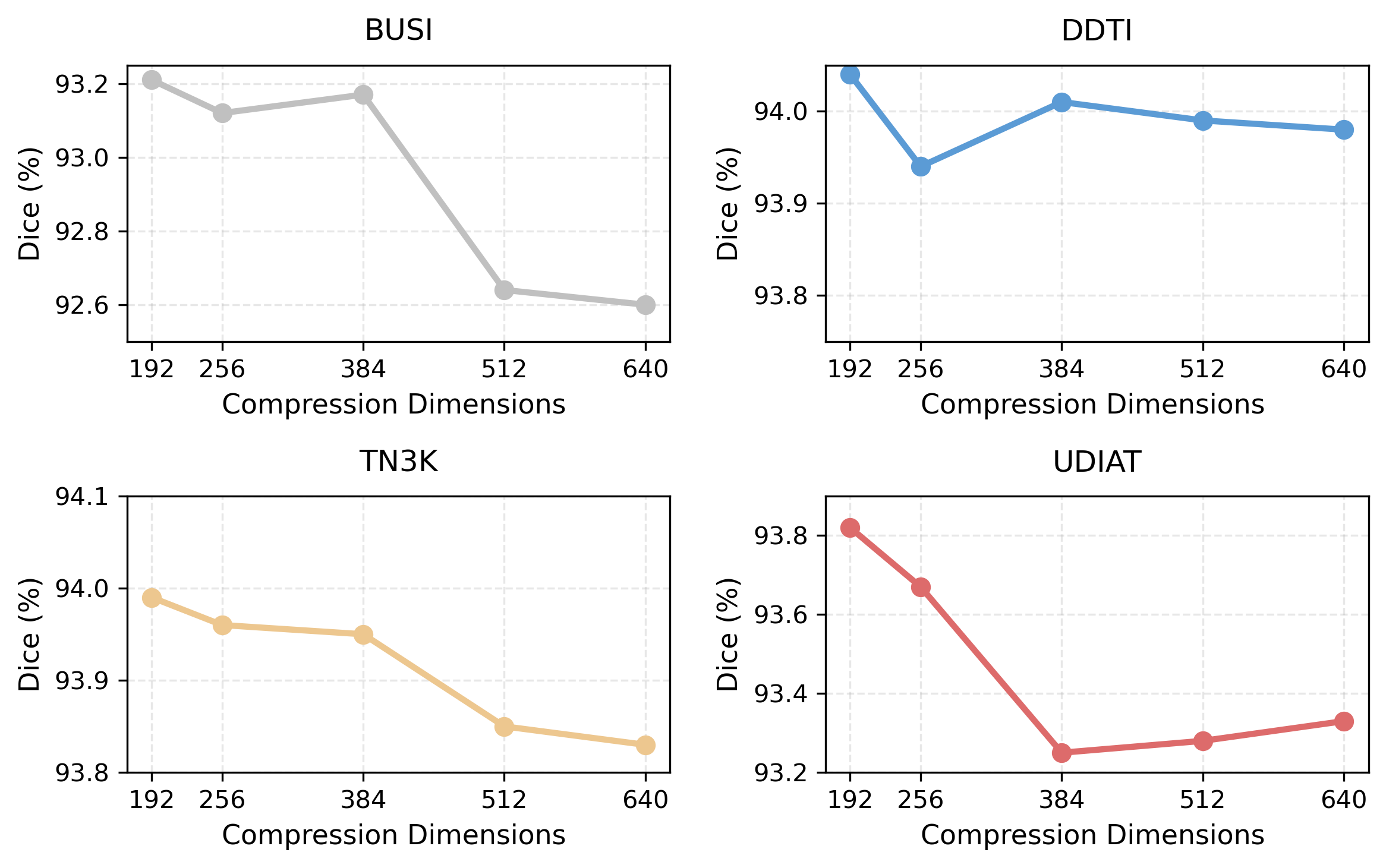}
  \caption{Ablation study of embedding dimension reduction in UniUltra-mini.}
  \label{fig:dimension}

\end{figure}

\begin{figure}[!t]
  \centering
  \includegraphics[width=1\linewidth]{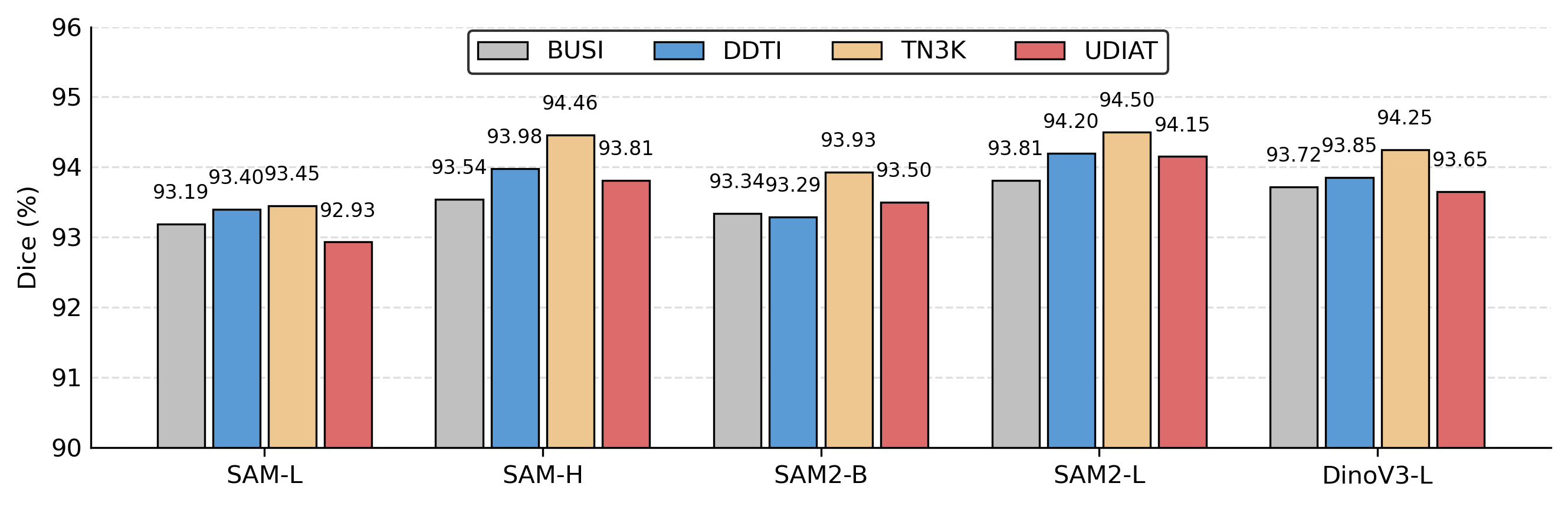}
  \caption{Impact of ViT backbone on UniUltra-L with different architectures and pre-trained weights.}
  \label{fig:encoder_comparison_styled}

\end{figure}

\subsection{Ablation study}
To further illustrate the necessity of each component, we perform ablation experiments where CH-Adapter and edge-aware ablations use the same datasets (BUSI, DDTI, TN3K, and UDIAT) as our internal validation. We aggregate the training sets from all four datasets into a unified training set, ensuring comprehensive representation of diverse ultrasound imaging scenarios. The reported ablation results are computed as the average performance across the respective test sets of each dataset used. DSKD ablation uses the same two datasets as our external validation. All the best-performing results are averaged based on the datasets used.

\subsubsection{Effectiveness of CH-Adapter}
As illustrated in Table \ref{tab:3}, we utilize the SAM2 with fine-tuned decoder as the baseline. It can be observed that all PEFT models are better than the baseline. However, our UniUltra displays the best performance, achieving further improvements over them. Specifically, compared to the second-best performing model (Adapter + LoRA), our proposed model showcases a 0.66\% increase in Dice, and a 9.41\% enhancement in HD performance. More importantly, our model requires fewer trainable parameters. These comparisons confirm the critical role of our proposed CH-Adapter, especially in achieving superior performance with fewer trainable parameters. It indicates that the proposed adapter architecture effectively leverages the pre-trained knowledge while introducing minimal computational overhead for domain-specific adaptation.

\subsubsection{Ablation Study of Distillation Level Strategy}
To investigate the contribution of each level in our hierarchical distillation strategy, we conduct ablation studies on different distillation level combinations, as shown in Table \ref{tab:level_distill}. The results clearly demonstrate that the three-level distillation strategy consistently achieves the best performance across all datasets and metrics. Compared to using only the third level or combining the second and third levels, incorporating all three distillation levels yields substantial improvements in both Dice score and HD. This validates that each distillation level captures complementary information from different feature hierarchies, and their combination is essential for optimal segmentation performance.

\subsubsection{Significance of Four Directional Edge-Aware Fillers} To investigate the efficiency of our edge-aware component, we design ablation experiments by incrementally adding Sobel filters, as illustrated in Table \ref{tab:4}. The CH-Adapter without any directional Sobel filters serves as the ablation baseline. By sequentially adding directional Sobel filters (horizontal, vertical, right diagonal, and left diagonal), the performance of the model gains progressive improvement. When all four directional filters are applied, the model displays the best performance. Compared to the baseline, it achieves a 0.79\% increase in Dice and a 13.92\% improvement in HD performance. These results highlight that adopting multi-directional edge information can effectively enhance model performance.

\subsubsection{Ablation Study of DSKD with external validation} 
Table \ref{tab:5} proves the effectiveness of our proposed Deep-Supervised Knowledge Distillation (DSKD). Referring to Equation \ref{eq:5} and Figure \ref{fig:2}b, the first configuration applies distillation on the integration ($D_1$) of Hiera block and CH-Adapter. The second configuration introduces additional distillation on the Hiera block ($D_2$). Similarly, the third configuration distills both the integration and CH-Adapter ($D_3$). By combining all three distillation positions, the complete three-layer distillation mechanism achieves the best results with a Dice score of 93.77\%, mIoU of 88.38\%, and HD of 42.53. Compared to only applying distillation on the integration ($D_1$), this comprehensive mechanism improves the Dice score by 1.42\% and enhances the HD performance by 27.76\%. These ablation experiments systematically evaluate the contribution of each distillation position and demonstrate the effectiveness of DSKD in improving generalization capabilities of UniUltra.

\subsubsection{Ablation Study of Embedding Dimension Reduction}
As shown in Figure \ref{fig:dimension}, for the embedding dimension reduction from 640 to 192, we conduct an ablation study by gradually reducing the dimension. Our findings indicate that 192 dimensions achieve the best overall performance across different datasets while significantly reducing computational cost. This suggests that 192 dimensions provide an optimal balance between model efficiency and segmentation accuracy, making it the ideal choice for our lightweight student model designed for real-time clinical applications.

\subsubsection{Ablation Study of Encoder Styles}
To validate our choice of encoder architecture, we conduct a comprehensive ablation study comparing various Vision Transformer (ViT) types and sizes, as shown in Figure \ref{fig:encoder_comparison_styled}. The results demonstrate that SAM2-L, which employs the Hiera-L architecture, achieves the best overall performance across multiple datasets. These results validate our design choice of using Hiera-L as the encoder backbone, as it offers an optimal balance between segmentation accuracy and computational efficiency.

\subsubsection{Ablation Study of Edge Detection Filters}
To demonstrate the effectiveness of choosing the Sobel filter, we conduct an ablation study comparing different edge detection filters. As shown in Table \ref{tab:edge_operator_comparison}, the Sobel operator achieves the best performance on most datasets, outperforming both Laplace and Canny operators with superior performance across BUSI, DDTI, and UDIAT datasets, while maintaining competitive results on TN3K. The consistent superiority of Sobel across multiple evaluation metrics validates its effectiveness in capturing edge information for ultrasound segmentation.

\subsubsection{Ablation Study of Prompts}
To comprehensively assess UniUltra's adaptability and robustness under different prompt configurations, we systematically compared different prompt methods across four internal validation datasets. As shown in Table \ref{tab:prompt_comparison}, the combination of 1 point and 1 bounding box achieves the best performance across all metrics, demonstrating superior segmentation accuracy and boundary precision compared to single-prompt strategies.

\section{Conclusion}
In this paper, we introduce UniUltra, a novel universal ultrasound image segmentation framework addressing domain adaptation and computational efficiency challenges. Our CH-Adapter consists of context-aware and edge-aware components. The context-aware component bridges the domain gap between natural and medical images, while the edge-aware component employs multi-directional Sobel filtering to enhance boundary detection. Moreover, we devise DSKD to achieve model compression while preserving robust feature extraction capabilities. Extensive validation on six public datasets demonstrates superior performance across diverse ultrasound scenarios. UniUltra achieves high segmentation accuracy with fewer parameters and reduced computational overhead, highlighting promising prospects for clinical deployment.

\bibliographystyle{IEEEtran}
\bibliography{refs}

\vfill

\end{document}